\documentstyle[pic]{article}

\let\citep\cite

\begin{document}

\title{Phenomenological Quantum Gravity
}


\author{Sabine Hossenfelder and Lee Smolin}
\address{Perimeter Institute for Theoretical Physics
 31 Caroline St. N, Waterloo Ontario, N2L 2Y5, Canada}

\maketitle


If the history of science has taught us anything,  it's that persistence and creativity makes the once impossible possible. It has long been thought experimental tests of quantum gravity are impossible. But during the last decade, several different approaches have been proposed that allow us to test, if not the fundamental theory of quantum gravity itself, so at least characteristic features this theory can have. For the first time we can probe experimentally domains in which quantum physics and gravity cohabit, in spite of our failure so far to make a convincing marriage of them on a theoretical level.  

Gravity is a very weak interaction. The only reason why we notice it so prominently in our day-to-day lives is that, unlike the other interactions, it cannot be neutralized. For all feasible Earth-based experiments examining short-distance physics, the gravitational interaction is completely negligible. For the same reason, experimentally testing quantum effects of gravity is hard. The effects are expected to become comparable to those of the other interactions only at energy scales close by the Planck energy, $E_{Pl}=\sqrt{\hbar c^5 / G}$. At $10^{16}$ TeV this energy is out of reach for collider experiments, and it is even far above the highest energies observed in cosmic rays. 

But nonetheless, the effects of quantum gravity can, in a few circumstances,  become observable.  A wide range of possible tests of hypotheses about quantum gravity have been proposed, and a number have even been carried out.  Most of these can be understood as testing a limit of quantum gravity where effects of quantum theory alone, or gravity alone can be ignored. Still, there may be quantum gravitational
phenomena since we can take $\hbar \rightarrow 0$ and $G \rightarrow 0$, while keeping their ratio,
$E_{Pl}=\sqrt{\hbar c^5/ G}$ fixed.  These are then phenomena governed by two parameters, $c$ and $E_{Pl}$.  

{\bf Tests of the symmetry of spacetime}

Most of the tests of hypotheses about quantum gravity in this regime concern the symmetries of spacetime which are assumed in particle physics.  Indeed, the most fundamental question one can ask about a physical system is what is the symmetry of its ground state.  We know that in classical physics, 
the groundstate is Lorentz-invariant
and the principles of special relativity are satisfied.  It is interesting to ask whether the same is the case when the effects of quantum gravity are considered.  Experiments are currently probing whether the Lorentz symmetry is preserved when effects of the order of the ratio of energies in the experiment to $E_{Pl}$ are taken into account.

One plausible hypothesis is that the principle of relativity breaks down at the scale $E_{Pl}$, so there is a preferred state of motion and rest.  Those who suggest this point to the existence of a preferred cosmological rest frame.   There are now quite good limits on this possibility.  Several come from the fact that if special relativity were false the speed of light would be no longer an invariant.  So one can look for a variation in the speed of light proportional to $E/E_{Pl}$, where $E$ is the
energy of a photon.  That is one looks for an energy dependent speed of light of the form $v=c(1\pm a E/E_{Pl} )$, where $a$ is a dimensionless parameter to be determined.  

This effect can be looked for in light coming great distances from astrophysical sources such as gamma ray bursts.  Even if the effect is tiny, these gamma ray bursts are billions of light years away and the arrival time of a photon can be offset by a few seconds.  This has not been seen, so we know that the parameter $a$ must be less than around one\cite{Fermi-nature-bound} (at least for the minus sign).  

The Fermi telescope, which was launched in June 2008, has detected a much larger number of very high energetic gamma ray bursts than previously expected. And the arrival of photons from these bursts also offered surprises. Several bursts have now been documented in which the higher energetic photons ($>$GeV) arrive with a delay of more than 10 seconds after the onset of the burst has been observed in the low energy range (keV-MeV). While it is still unclear whether this delay of the high energetic photons is caused at emission or during propagation,  more statistics and a better analysis –- in particular about the delay's dependence on the distance to the source -- will eventually allow to narrow down the possible causes and constrain models that give rise to such features \cite{AmelinoCamelia:2009pg}.

Models with a breaking of Lorentz symmetry indicate that the effect should depend on polarization, so that the sign would be plus for one polarization and minus for the opposite.  The result is that the planes of polarization rotate as the photons travel, in a way that leads to polarized light becoming unpolarized.  From the fact that we see polarized light coming from distant galaxies and nebula it has been shown that we must take $a < 10^{-9}$ \cite{crab-bound}.  

Another test of the principle of relativity is a prediction that very high energy cosmic ray protons interact with the cosmic microwave background (CMB).  Given just the principles of special relativity these interactions were predicted to take place at an energy of above $10^{19}$~eV, and the result is that the protons lose energy. This leads to what is called the GZK cutoff on the cosmic ray spectrum, which says that we should not see cosmic ray protons coming from further away than a certain distance, $75$ megaparsecs, which is the mean free path for this interaction.  This prediction was confirmed recently by observations at the AUGER cosmic ray detector \cite{AUGER}. 

These and other results make it seem very unlikely the principle of relativity breaks down at Planck scales, at least at order $E/E_{Pl}$.  The results to date do however allow a more subtle hypothesis called deformed special relativity, in which the principle of relativity is preserved, but in a way as to make all observers agree about what the special energy $E_{Pl}$ is \cite{DSR,Magueijo:2001cr}.  It is also possible that special relativity breaks down, but only in a way that can be seen by experiments sensitive to effects of the order of $(E/E_{Pl})^2 $.  But the good news is that given the rate of progress of this field, these more subtle hypotheses could be tested and, if not true, be ruled out, within a decade.

Another very interesting possibility is that quantum gravity effects fail to be symmetric under the discrete transformations of physics, such as parity (P) or (CP) or time reversal invariance.  These might show up in high precision measurements of these effects.   

The hypothesis that quantum gravitational effects break parity symmetry also has implications for observations of the CMB.  It leads to predictions for a signal in the CMB spectrum, which would show up in correlations between the temperature fluctuations and certain parity odd polarization modes, called $B$-modes \cite{TB}.  So far such an effect has not been observed, but the Planck satellite is expected to probe this effect much more sensitively than before.  

{\bf Brownian motion and stochastic effects}

A common hypothesis about quantum spacetime is that space and/or time become discrete on the Planck scale, in much the same way as matter becomes discrete when examined at the scales where atoms can be perceived.  It is interesting to recall that the atomic hypothesis was confirmed first, long before atoms were seen directly, by the observations of effects from their random motion.  This was the great work by Einstein on Brownian motion in 1905.  Similarly, it is possible to imagine that the effects of a fundamental discreteness of space and/or time would show up in random fluctuations on the propagation of light or elementary particles.  There can also be other motivations besides discreteness for stochastic effects, such as the conjecture that quantum fluctuations in the geometry of spacetime cause the light cones to fluctuate, thus affecting the speed of photons in a random way \cite{lightconefluc}.  

Such effects would show up as noise in the incredibly sensitive interferomters that are used to measure gravitational radiation \cite{noise-AC,Hogan}.  Remarkably, under some simple hypotheses, the modern gravitational wave detectors have sensitivity to this at order $E/E_{Pl}$.  Presently there seems to be no sources of noise seen in the detectors that is not accounted for by more mundane explanation, so the evidence is that there are no such effects at this order.

{\bf Quantum effects in strong gravitational fields}

Now we come to effects which require quantum effects themselves, so that we cannot neglect $\hbar$ and $G$.  The first of these to be predicted is of course Hawking radiation, and it is tantalizing that were primordial black holes created in the early universe with masses of around $10^{15}$~grams (or a good-sized mountain) the last stages of their Hawking radiation would be currently visible as bursts of $x$-rays.   It is disappointing that despite searches, no such bursts have been seen because, if they were, the precise spectra would provide tests of theories of quantum gravity, such as loop quantum gravity \cite{LQG-BHspectra}.

Even if evaporating black holes are not observed, a fascinating possibility which has been proposed is to construct analogues in condensed matter systems, of evaporating black holes. 
These would not test quantum gravity directly, but they would provide tests of the reasoning which leads to the prediction of Hawking radiation.

The next  obvious place to look is the early universe, where gravity must have been very strong. Quantum gravity could have left traces in the CMB by quantum effects affecting the dynamics of the universe's expansion. This data could contain information about quantum corrections to the evolution equation, or even a possible phase transition from a pre-geometrical to a geometrical phase \cite{Konopka:2006hu,CJL}.

Beyond that a big challenge for quantum gravity theories is an understanding of the big bang itself.  Was it really the first moment of time? In this case a quantum gravity theory needs to provide initial conditions for the universe and this might, possibly, imply some predictions for cosmological observations.  Or is the cosmological initial singularity predicted by general relativity only an artifact of the neglect of quantum physics, and is it really replaced by a bounce or a transition from an earlier universe \cite{bounce}.  A very exciting possibility is that such a bounce-and not inflation-would be the right explanation for the observed fluctuations in the CMB.  If so then those observations are seeing quantum gravity effects.

{\bf The hypothesis of large extra dimensions}

A completely different category of models studies the possibility that quantum gravitational effects could be much stronger than usually thought due to a modification of the gravitational interaction on shortest distances. Such a modification occurs in scenarios with large additonal spatial dimensions 
whose existence is predicted by string theory, 
and has the consequence that quantum gravity could become observable in Earth-based collider experiments, such as the Large Hadron Collider (LHC) that will start operation soon. 

If this should turn out to be a correct description of Nature, we would see the production of gravitons and black holes at the LHC \cite{Landsberg:2008ax}. The gravitons themselves would not be captured in the detector and lead to a missing energy signal, the missing particles having spin 2. Black holes would decay via Hawking radiation. Ideally the distribution of decay products would allow to determine the parameters of the model, the number and size of the extra dimensions. Black hole production and decay would be a striking signature, and allow us to examine the fate of black hole information during the evaporation process in the laboratory.

{\bf Outlook}

It should be emphasized that the experiments we have discussed are described by phenomenological models that are, at least so far, not derived from any of the presently pursued approaches towards quantum gravity. The purpose of these models is to study consequences that arise from specific features the underlying theory could have, and ideally constrain them. In such a way, we could learn about the general properties of the theory we are trying to find, for example whether it does have additional spacelike dimensions, or results in a deviation from Lorentz invariance. 

Nonetheless, it is remarkable that in the last few years the precision of tests of hypotheses about quantum gravity has increased dramatically, to the point that these experiments regularly probe effects at the Planck scale and beyond.  

Progress in physics needs two ingredients: mathematical consistency and experimental evidence. Relying entirely on mathematical consistency is a shot into the dark. It comes with the burden of connecting a new theory, and possibly a completely new mathematical framework, back to what we already know. Experimental evidence sheds light into the darkness, and helps to narrow the range of possible theories. Theory and experiment work best together. The increase of attention and effort that the phenomenology of quantum gravity has seen within the last decades is thus a very welcome and long overdue contribution to our quest of finding a unifying description for all interactions of the standard model.

\end{document}